\documentclass[aps,prd,twocolumn,groupedaddress,nofootinbib]{revtex4}

\usepackage{graphicx}
\def\beq{\begin{equation}}
\def\eeq{\end{equation}}
\begin{document}
\title{Lorentz violation and perpetual motion}

\author{Christopher Eling}
\author{Brendan Z. Foster}
\author{Ted Jacobson}
\author{Aron C. Wall}
%
\affiliation{Department of Physics, University of Maryland, College
Park, MD, 20742}

\date{\today}

\begin{abstract}

We show that any Lorentz violating theory with two or more
propagation speeds is in conflict with
the generalized second law of black hole thermodynamics.
We do this by identifying a classical energy-extraction method,
analogous to the Penrose process, which would decrease the
black hole entropy. Although the usual definitions of black hole
entropy are ambiguous in this context, we require
only very mild assumptions about its
dependence on the mass.
This extends the result found
by Dubovsky and Sibiryakov, which uses the Hawking  effect
and applies only if the fields with
different propagation speeds interact just through gravity.
We also point out instabilities that could interfere with
their black hole {\it perpetuum mobile}, but argue that these can be
neglected if the black hole mass is sufficiently large.

\end{abstract}


\maketitle
%
%
\section{Introduction}
Is Lorentz symmetry an exact symmetry in nature, or is it only
approximate? In order to address this question, several models have
been proposed in which some dynamical fields break Lorentz
symmetry.
In such
models different fields can have different maximal
speeds of vacuum propagation, as measured in a preferred reference
frame at each point.
When gravity is included,
black hole solutions can
exist~\cite{Giannios:2005es,Mukohyama:2005rw,
Babichev:2006vx,Eling:2006ec}
with
multiple, nested horizons, one
for
each maximal speed
of propagation in the theory.
Each horizon traps the corresponding species of field excitations.
Only
the innermost horizon is a true event horizon, which traps all
information inside of it.

Are
the laws of thermodynamics
obeyed by
black hole systems in Lorentz-violating theories, as they are in
standard General Relativity? An
ominous
new
feature is that the
multiple horizons will generally have different surface gravities
and therefore different temperatures.
This
conflicts with the ``zeroth law", by which a system in thermal
equilibrium has a single temperature.
Consequently, the entropy cannot be determined via
the usual relation $dS = dE/T$.
Also, the usual identification of the entropy with horizon
area becomes ambiguous since there are multiple
horizons. Moreover, the entropy might not even be
proportional to {\it any} area.
Perhaps related to
these problems
is the failure of
the
Noether charge algorithm~\cite{Wald:1993nt} for identifying the
entropy of a stationary black hole when applied to Einstein-aether
theory~\cite{Foster:2005fr} and, by extension, other theories of
gravity with a dynamical preferred frame.

Nevertheless, Dubovsky and Sibiryakov (DS)~\cite{Dubovsky:2006vk}
were able to investigate the status of the second law in
a Lorentz-violating gravity theory by considering a process in
which the macroscopic state of the black hole is held fixed.
Their analysis is presented
in the context of the ghost condensate
theory~\cite{Arkani-Hamed:2003uy}
but, as they suggest, it should apply more generally to
any Lorentz-violating gravity theory with multiple maximal speeds.
DS describe a {\it perpetuum mobile} that
pumps heat from a colder to a hotter reservoir,
by taking advantage of the Hawking effect and
the differing temperatures of the nested horizons.
They consider a static black hole, and
two fields $A$ and $B$
that travel at different maximal speeds $c_A$ and $c_B\sim c_A$,
with $c_{B}>c_{A}$. (We will use units in which $c_{A,B}\sim1$.)
Via the Hawking effect, the $A$ and $B$ horizons
thermally radiate the corresponding species of particle, with
$T_B > T_A$ since the Hawking temperature scales
inversely with the horizon radius in the ghost condensate theory.
DS assume the $A$ and $B$ fields have no interaction
except through gravity.
To construct the
device,
they place $A$ and $B$ shells surrounding the
black hole that interact only with $A$ and $B$ fields respectively. They
then show that it is possible to choose the temperatures of the
shells such that
\beq
    T_{B,Hawking}>\ T_{B,Shell}
    >T_{A,Shell}> T_{A,Hawking}
\eeq
and such that the energy fluxes balance one another so that the
black hole stays the same size.
Energy flows
from the colder $A$ shell into the black hole, and from the black
hole to the hotter $B$ shell.
The second law thus appears to be violated.

DS consider three possible
ways this conclusion might be evaded~\cite{Dubovsky:2006vk}.
We quote:
\begin{enumerate}
    \item[``(i)] The presented description of the Hawking radiation
    in the ghost condensate is correct, but there is some subtle way in which a
    low energy effective theory forces our perpetuum mobile to change its state
    so that the entropy actually increases.
     \item[(ii)]  The derivation of the Hawking radiation using only low energy
    theory is incorrect.
    \item[(iii)] The presented description of the Hawking radiation in the ghost
    condensate is correct, and the violation of the second law of thermodynamics
    within a low energy effective theory is a physical effect. According to the discussion in  the
    Introduction this
    means that the UV completion of the ghost condensate, if it exists at all,
    has very unusual properties."
\end{enumerate}
DS put forth some arguments against the first two possibilities, but
do not claim to have ruled them out conclusively.
Our interpretation of what DS mean by (iii) (considering related remarks made in their paper)
is as follows: the device works
as described,
but another version of the second law remains valid if, due to
nonlocality or unbounded propagation speed in the UV completion,
the notion of a causally hidden black hole region is eliminated.
In this case, the
    entropy increase inside the black hole can influence the outside
    state, and the total entropy, inside and out, is nondecreasing.

 This last point highlights the need to distinguish two different
versions of the ``second law": the ordinary second law (OSL) and
the generalized second law (GSL)~\cite{Bekenstein:1973ur}.
The OSL refers to the total
entropy, as counted both inside and outside black holes,
whereas the GSL replaces the inside entropy by a special
``black hole entropy" $S_{\rm bh}$ determined by the macroscopic geometry alone.
The validity of the OSL for quantum fields in curved spacetime
on a complete spacelike foliation
is unaffected by the presence of black holes. It should thus be
valid in Lorentz-violating theories. Therefore exotic properties of
a UV completion are not required
to uphold
the OSL.
Only the validity of the GSL is in question.

The GSL states that the generalized entropy
$S_{\rm bh}+ S_{\rm outside}$
cannot decrease. It is not obvious here which region
is the ``outside" for defining  $S_{\rm outside}$.
One might suppose it should be
the outside of the innermost causal horizon,
but for the purposes of this paper it will not be necessary
to  specify exactly which region is the ``outside".
In General Relativity, $S_{\rm bh}$ is one quarter the horizon area
in Planck units. In the Lorentz violating case, DS did not need
to specify $S_{\rm bh}$, since the macroscopic properties of their
black hole were held fixed. In section \ref{Classical} we shall make only
some weak assumptions about the form of $S_{\rm bh}$.

 \section{Summary of our results}

It is surprising that a thought experiment with black
holes could reveal such an unexpected and drastic consequence of Lorentz
violation. We thus set out to find a flaw in the proposed
{\it perpetuum mobile}. However, rather than finding a flaw, we found only further
support for the GSL violation.

We first consider two processes not discussed by DS that could
potentially destabilize the device: (i) gravitationally
mediated equilibration of $A$ and $B$ species in each shell, and
(ii) classical or quantum instability of the ergoregion between the
$A$ and $B$ horizons. We shall argue that none of these phenomena
can save the GSL in all circumstances.

Next we present a {\it classical} process by which energy can be
extracted from the black hole much quicker than any instability we
are aware of, lowering the black hole entropy and thus violating the
GSL. This process sidesteps the use of Hawking radiation, and
permits direct interaction between the $A$ and $B$ fields. It should
also be possible to violate the GSL by dumping heat into the black
hole and then using this classical process to extract the
corresponding energy without entropy, thus lowering the outside
entropy without a net change of the black hole
entropy.

That the microstates of the heat energy cannot engender 
gravitational ripples outside with equal entropy has generally been
assumed in discussions of black hole thermodynamics.
This assumption is plausible and might be established using
a multipole expansion of the source, together with assumed
quantization of graviton number. We assume it without proof here,
along with the extension to include perturbations of the Lorentz violating fields.

Throughout this paper we assume that the black hole mass and
radius are related by $R\sim GM$, as in General Relativity,
the ghost condensate theory~\cite{Mukohyama:2005rw} , and
Einstein-aether theory~\cite{Eling:2006ec}.

\section{Destabilizing processes}

In this section we discuss the rate of various
processes that could potentially destabilize the black hole
{\it perpetuum mobile}.
We will
argue that they can be ignored
for systems in which
the gravitational coupling is sufficiently
weak.

\subsection{Equilibration of species}

It was stipulated by DS that the $A$ and $B$ fields do not interact
directly with
each other.
Since, however,
they both interact with gravity, they must at least have
gravitationally mediated interactions. This implies that in true
equilibrium the $A$ and $B$ species must be thermally populated in
each shell.
But then the device malfunctions,
since the colder $A$ shell
absorbs
heat from the hotter $B$ shell and $B$ horizon.
Nevertheless, it
could operate for long
enough to violate the
GSL, if the equilibration were slow enough compared to
the heat pump rate.

Rather than attempt here to estimate the actual equilibration rate,
we instead employ a simple scaling argument. The gravity mediated
equilibration rate
can be
decreased by
``turning down" the gravitational constant.
Meanwhile
the heat pump rate can be held fixed by scaling $M$ so that $R\sim GM$,
and therefore the Hawking temperatures and absorption and emission
cross sections, remain fixed. Thus, for sufficiently weak
gravitational coupling, the
GSL
can be violated before
equilibration ensues.

Since $G$ is not dimensionless, turning it down must be equivalent to
leaving it fixed while scaling the system parameters.
If we replace $R$ by $\lambda R$, and divide the shell temperatures
by $\lambda$ to match the scaling of the Hawking temperatures,
the DS entropy pump rate will scale as $1/\lambda$, since it then depends
only on unchanged dimensionless parameters and the radius $\lambda R$.
On the other hand, the entropy production due to gravity-mediated species
equilibration scales with an additional factor of $1/\lambda^4$.
This is because the gravitational coupling between two particles
scales with the particle energies, which in turn each scale
with the temperature as $1/\lambda$, and the amplitude is
squared to obtain the rate.
So by increasing
$R$ and decreasing the shell temperatures
the equilibration rate
can be made much slower than the pump rate.

\subsection{Ergoregion instability}
An ergoregion is a place where the asymptotic time translation
Killing vector of a spacetime becomes spacelike, allowing negative
energy states to exist. The Hawking effect is an instability brought
about by the existence of an ergoregion hidden behind a horizon.
 But
if an ergoregion exists {\it outside} a horizon then other
instabilities can arise.  A rotating black hole, for example, exhibits
superradiant
scattering---the amplification of classically scattered
fields---and
therefore is unstable to
quantum spontaneous emission~\cite{zeldovich,starobinsky,unruh}.
Both effects result
in
a
transfer
of the body's rotational
energy to outgoing field modes.

Could
spontaneous ergoregion decay
occur also for
the {\it perpetuum mobile}?
Since the slower, $A$
field will possess an
ergoregion that lies outside the faster, $B$ horizon, processes can
occur in which negative energy $A$ particles that fall across the
inner horizon are generated along with positive energy $B$ particles
that escape to infinity. These could in principle compete with the
Hawking flux. Any such process, however, must be gravitationally
mediated if $A$ and $B$ particles do not interact directly.
Therefore, as argued above, the rate of these processes can be
suppressed below that of the Hawking flux by turning down the
gravitational constant. If instead direct $A$-$B$ interactions
exist with dimensionless coupling, then ergoregion decay would
scale with $R$ in the same way as does Hawking radiation,
potentially interfering with the {\it perpetuum mobile} unless the
coupling is sufficiently weak.

One might worry about
exponentially growing
instabilities. These are known to occur if the positive energy radiation
 returns coherently to the ergoregion, or the negative energy radiation
 remains in the ergoregion.  Either way, emission
of further radiation can be stimulated.
For example, if a rotating black hole is surrounded
by a mirror,
the outgoing positive energy modes can be reflected back to the ergoregion
creating  a
``black hole bomb"~\cite{bomb}.
(The same thing can happen with
the mirror replaced by
anti-de Sitter boundary conditions\cite{Cardoso:2004hs}.)
Alternatively, a rotating star with an ergoregion but no
horizon is unstable because the negative energy radiation
piles up in the ergoregion~\cite{stars}. The {\it perpetuum mobile} could
perhaps be similarly unstable due to a gravity-mediated process
in which negative energy $A$-modes are stimulated along with
positive energy $B$-modes which are coherently reflected off the $B$ shell.
However, to be unstable
the total amplitude for this process must exceed a critical value.
For sufficiently small gravitational coupling
or shell reflectivity, no instability will occur.

\section{Classical violations of the second law}
We now turn from the {\it perpetuum mobile} of DS to a purely classical
process that leads to GSL violation.
It
makes no use of the Hawking effect, vitiating
the need to verify
that effect in this Lorentz violating context.
Instead, it takes advantage of the
$A$
ergoregion
in a way analogous to the
Penrose process in
the ergoregion of a rotating black hole~\cite{Penrose:1971uk}.

\subsection{Mass and entropy}\label{Classical}
We begin by discussing the connection between extracting energy from
the black hole and lowering its entropy.
As discussed in the Introduction, it is not
yet clear how the
black hole entropy should be defined in a Lorentz
violating theory.
Therefore, we will  make only the following mild
assumptions about the entropy $S(M)$ of black holes of mass $M$
and size $R\sim GM$:
\begin{enumerate}
 \item When $M_1 \gg M_2$, then $S(M_1) \gg S(M_2)$.
  \item By choosing $M$ sufficiently large, the entropy of any radiation
    emitted by the hole over a time $R$ can be made to be an arbitrarily small
    fraction of $S(M)$.
\end{enumerate}

Note that  Hawking radiation, and
ergoregion decay with dimensionless $A$-$B$ interaction,
both produce entropy at a rate scaling
as $1/R$, since $R$ is the only relevant length scale.
To satisfy both assumptions, it is therefore sufficient that $S(M)$
increase with $M$ at least as fast as $M^\alpha$ for
some $\alpha>0$.

We will show that a process exists that reduces the energy of
a black hole by an amount proportional to $M$ over a time
of order $R$, without any
incidental entropy increase outside the black hole. By repeating this process
one can shrink the black hole down to a much smaller size in a time
proportional to $R$. The first of the above assumptions implies
the final black hole then has much smaller entropy, and the second assumption
implies that if one starts and ends with a sufficiently large black hole,
any radiated entropy is negligible. Thus the process violates the GSL.

\subsection{Classical energy extraction from black holes}

We now discuss the spacetime structure, which gives rise to the conservation
laws governing the energy extraction process.  Let $g_{ab}$ be the metric felt
by the $A$ field. The assumed Lorentz violation involves a
``preferred"  unit timelike vector $u^{a}$ that has unit norm with respect to $g_{ab}$:
\beq
    g_{ab}u^{a}u^{b}=1. 
\eeq
The metric
$\tilde{g}_{ab}$ felt by
the $B$ field
 is
given
(up to an arbitrary conformal factor)
by
\beq
    \tilde{g}_{ab}=u_{a}u_{b}+\frac{c_{A}^{2}}{c_{B}^{2}}(g_{ab}-u_{a}u_{b}),
\eeq
where the index on  $u_{a}$ is lowered using $g_{ab}$.
We
are considering
a
black hole
spacetime in which $g_{ab}$,  $\tilde{g}_{ab}$, and $u^a$ are
all spherically
symmetric and
static, with asymptotically timelike
Killing field $\xi^a$.
For each metric, $\xi^a$ is timelike outside and spacelike inside the
corresponding horizon.

The 4{}-momentum covector $p_{a}$ of a particle is naturally defined
in a metric{}-independent way as the gradient of the
Hamilton{}-Jacobi principal function associated with the particle.
This 4-momentum is locally conserved.
The Killing energy $\cal E$ is defined by ${\cal E}=p_a\xi^a$.
The mass shell conditions
depend on the metric; for example
massless $A$ particles satisfy $g^{ab}p_ap_b=0$ while massless $B$
particles satisfy $\tilde{g}^{ab}p_ap_b=0$, where
 $\tilde{g}^{ab}=u^{a}u^{b}+(c_{B}^{2}/c_{A}^{2})(g^{ab}-u^{a}u^{b})$
 is the inverse of $\tilde{g}_{ab}$.
The energy $E$ and 3-momentum $\vec{p}_a$ in the
preferred frame are defined by $p_a=Eu_a+\vec{p}_a$, where
$\vec{p}_au^a=0$.  Massless $A$ particles then satisfy
$E^2=g^{ab}\vec{p}_a\vec{p}_b$, while massless $B$ particles
satisfy  $E^2=(c_{B}^{2}/c_{A}^{2})g^{ab}\vec{p}_a\vec{p}_b$.
Hence the $B$ null covector cone lies {\it within} the $A$
null covector cone.

Now let a system $\Sigma$ composed of $A$ and $B$ particles fall through
the $A$ horizon,
meeting at a point $x$ in the $A$ ergoregion outside the $B$ horizon.
Henceforth we refer to this zone as simply the ``ergoregion".
We will assume that the energy in
$\Sigma$ is sufficiently small
that it does not appreciably disturb the black hole.
We also require $\Sigma$ to be well localized
compared to the size of the
ergoregion so that it can be treated as
a  ``point particle".
These conditions can both be satisfied
if the energy in the system is less than the black hole mass
times some fixed small constant $k$ that depends on the
particular theory. For example if $c_A$ and $c_B$ are very close,
then the ergoregion is very thin and $k$ must be correspondingly smaller.

\begin{figure}
    \includegraphics[width=80mm]{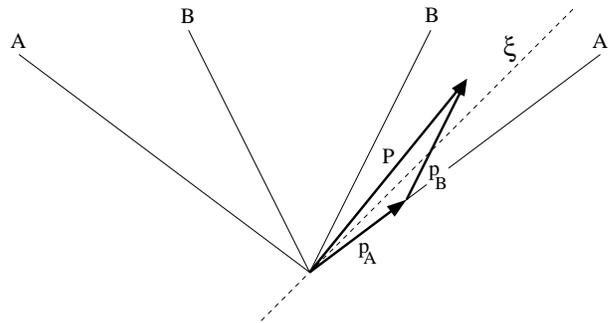}
    \caption{The radial 4-momentum covector space at the point $x$ in
    the ergoregion ($p_a m^a=0$ for all vectors $m^a$ tangent to the
    symmetry sphere through $x$).
    The momenta with vanishing Killing energy ${\cal E}=p_a\xi^a$
    lie along the dashed line. ${\cal E}$
is negative for $p_A$, since $x$ lies in the $A$ ergoregion,
while it is positive for all momenta on the $B$ null cone.
The total 4-momentum $P$ has positive ${\cal E}$, hence it points above
the dashed line.
\label{FIG}}
\end{figure}

We arrange $\Sigma$ so that
at the meeting point $x$ its
 net 4{}-momentum $P_a$
  is radial and outward pointing, lying outside the $B$-metric
momentum-space null cone as depicted in Figure~\ref{FIG}.
Further the system
should have positive Killing energy  so it can have
come from outside the outer horizon. A system containing just one
massive $A$ particle, for example, can satisfy these conditions,
if dropped in from just outside the $A$ horizon with 4-velocity sufficiently
close to the outgoing $A$-null ray.
However,
since we want to arrange for ejection of a $B$ particle in a classical
process we should start with at least one $B$ particle in the system.
One
possible scenario is that the $A$ and $B$ components fall in together in
a gravitationally bound configuration, or they could just be
arranged to meet in the ergoregion and interact there.
The net 4-momentum can still satisfy the required conditions
if the $A$ 4-momentum dominates.

After $\Sigma$  has fallen into the ergoregion, we imagine it
splits at $x$, where it has 4-momentum $P$,
into two separate components, one consisting of outgoing
massless $A$ particles with 4-momentum $p_A$,
and the other outgoing massless $B$ particles with 4-momentum $p_B$.
The total 4-momentum  covector can be conserved in such a process,
as illustrated in Figure~\ref{FIG}. The $A$'s then fall across the
$B$ horizon
carrying negative Killing energy, while the $B$'s escape outwards
across  the $A$ horizon. Since the Killing energy is conserved,
these carry out more energy than originally fell in, so the black hole
mass decreases. The mass decrease scales with the energy of
$\Sigma$, whose upper bound is $kM$, so the
mass can be decreased by some fraction $k'M$.

So far we have only imposed 4-momentum conservation; we have  not
addressed what kind of interaction could result in the
final $A$ component of the system having negative Killing energy. Since the
initial $A$ component falls into the ergoregion with positive Killing energy,
some Killing energy transfer to the $B$ component must be effected.
This transfer requires an interaction,
but $A$ and $B$ always interact at least gravitationally.
Seeing as the conservation laws permit the process, and an interaction
capable of mediating it exists, we shall presume that it can be
achieved. The black hole size $R$ sets the timescale
for the process, since the particles need only travel this distance,
and the energy transfer occurs on a much smaller length scale.

The outgoing $B$ particles need not carry any entropy at all. This is because
the system may be prepared in a pure state, and the whole
splitting process can occur via classical deterministic evolution, which
does not generate any entanglement entropy between the $A$ and $B$
components. This process therefore reduces the mass of the black
hole without creating any compensating matter entropy outside of the
black hole. Given our assumptions above,
by repeating this process many times, the GSL can be violated.

\section{Discussion}

We have identified possible destabilizing mechanisms that might have
interfered with the black hole {\it perpetuum mobile} devised by
Dubovsky and Sibiryakov, but found that they can be neglected for
sufficiently large black holes. Furthermore, we devised a classical
energy extraction process, which strengthens the case for GSL
violation in Lorentz violating theories. Unlike the DS {\it
perpetuum mobile}, it does not rely on the Hawking effect, and the
entropy decrease occurs much more quickly. Most importantly, it can
operate even if the $A$ and $B$ species have direct interactions.
Thus the GSL violation is shown to occur for a much broader class of
Lorentz violating theories with multiple speeds, not only those with
limited interactions.

Is this violation of the GSL necessarily unacceptable?
It would certainly seem unlikely to
 have led to any
observable consequences, given our current state of astrophysical
observation.
Moreover,
there is no \textit{a priori} reason why the
GSL
should hold,
considering the fact that the outside of a black hole is not a
closed system. From this perspective it is
perhaps
more surprising that the
GSL
holds for Lorentz symmetric systems, than that it
might fail for the rest.

On the other hand, if the GSL does {\it not} hold, then the
apparently deep connection between black holes and thermodynamics
would have been a coincidental false lead, not arising from
fundamental principles. This is true even if the difference in
speeds were very small so it would take a very long time to execute
a violation of the GSL.

GSL violation might be avoided if the UV completion
of the Lorentz violating theory eliminates the notion of a causally
hidden black hole region. Then the ``outside" would
include the black hole interior, leaving no ``black hole" contribution
to the generalized entropy, thus reducing the GSL to the OSL.
But this eliminates the essence of black hole thermodynamics.
So it appears that the only way to save black hole
thermodynamics is to reject the sort of Lorentz violation
considered here (and likely any other sort involving
Lorentz violating dispersion).

In retrospect, it is perhaps not so mysterious that the validity of
black hole thermodynamics is tied to Lorentz symmetry. After all, at
the root of the thermality of the Hawking effect lies the Unruh
effect: the vacuum is a thermal state with respect to the boost
Hamiltonian when restricted to the Rindler wedge~\cite{Sewell}. This
in turn can only hold for the vacuum of interacting fields if they
share a common Lorentz symmetry. \vspace{-4mm}
\begin{acknowledgments}
This work was supported in part by the National Science Foundation
under grants PHY-0300710 and PHY-0601800 through the University of
Maryland and PHY99-07949 through the KITP, Santa Barbara.
\end{acknowledgments}
%
%

%
%
\end{document}